\documentclass[fleqn,11pt]{article}
\usepackage{amsmath,amsmath,amsthm,amssymb,subfigure,outlines,epsfig,mycommands1}
\usepackage[small]{caption}

\usepackage[round]{natbib}
\usepackage[margin=1in]{geometry}

\usepackage{hyperref} 
\usepackage{stackrel}
\usepackage{flafter} 
\usepackage{setspace}
\usepackage{chngcntr}
\usepackage{apptools}
\AtAppendix{\counterwithin{Theorem}{section}}
\numberwithin{equation}{section}

\begin{document}

\newtheorem{Theorem}{Theorem}[section]
\newtheorem{Lemma}[Theorem]{Lemma}

\title{Nonconvex penalized multitask regression using data depth-based penalties}
\date{}
\author{Subhabrata Majumdar, Snigdhansu Chatterjee}
\maketitle

\begin{abstract}
We propose a new class of nonconvex penalty functions, based on data depth functions, for multitask sparse penalized regression. These penalties quantify the relative position of rows of the coefficient matrix from a fixed distribution centered at the origin. We derive the theoretical properties of an approximate one-step sparse estimator of the coefficient matrix using local linear approximation of the penalty function, and provide algorithm for its computation. For orthogonal design and independent responses, the resulting thresholding rule enjoys near-minimax optimal risk performance, similar to the adaptive lasso \citep{Zou06}. A simulation study and real data analysis demonstrate its effectiveness compared to some of the present methods that provide sparse solutions in multivariate regression.
\end{abstract}


\newpage

\section{Introduction}

Consider the multitask linear regression model:
$$ Y = X B + E $$
where $Y \in \mathbb R^{n\times q}$ is the matrix of responses, and $E$ is $n\times q$ the noise matrix: each row of which is drawn from $\mathcal{N}_q (0, \Sigma)$ for a $q \times q$ positive definite matrix $\Sigma$. We are interested in sparse estimates of the coefficient matrix $B \in \mathbb R^{p \times q}$, which are useful for inference in regression problems with a large number of predictors that have differential influences on multiple correlated response variables: for example in gene-expression data \citep{LozanoSwirszcz12, MolstadRothman16} and prediction of stock returns \citep{RothmanEtal10}. This is done through solving penalized regression problems of the form
\begin{equation}\label{eqn:penEqn}
\min_B \Tr \{ (Y - XB)^T ( Y - XB) \} + P_\lambda(B)
\end{equation}

The frequently studied single-response linear model may be realized as a special case of this with $q = 1$. In this setup, obtaining sparse estimates of the coefficient vector $\beta$ involves solving an optimization problem with the penalty function $P(\beta) = \sum_{j=1}^p p(|\beta_j|)$:
\begin{equation}\label{eqn:eqn01}
\hat \beta_n = \argmin_\beta \left[ \sum_{i=1}^n \rho ( y_i - X_i^T \beta) + \lambda_n \sum_{j=1}^p p(|\beta_j|) \right]
\end{equation}
for a general loss function $\rho(.)$, with $\lambda_n$ being a tuning parameter depending on sample size. The penalty term is generally a measure of model complexity that controls for overfitting. Starting from LASSO \citep{Tibshirani96} which uses the $l_1$ norm, i.e. $p(z)=|z|$, relevant methods in this domain include adaptive LASSO \citep{Zou06} that reweights the coordinate-wise LASSO penalties based on the Ordinary Least Square (OLS) estimate of $\beta$, and non-convex penalties proposed by \cite{FanLi01} and \cite{CHZhang10} that limit influence of large entries in the coefficient vector $\beta$, resulting in improved estimation of $\beta$. Further, \cite{ZouLi08} and \cite{WangKimLi13} provided efficient algorithms for computing solutions to the nonconvex penalized problems.

For multiple responses, \cite{RothmanEtal10} showed that penalizing at the coefficient matrix-level results in better estimation and prediction performance compared to performing $q$ separate LASSO regressions. Here the coefficient matrix $B$ has two levels of sparsity. The first level is recovering the set of predictors having non-zero effects on all the responses, while the second level of sparsity is concerned with recovering non-zero elements \textit{within} the non-zero rows obtained from the first step. Previous studies have performed this using either a bi-level penalty function \citep{VincentHansen14, LiNanZhu15} or a group lasso penalization to recover non-zero rows followed by within-row thresholding \citep{ObozinskiEtal11}. 

In this paper, we introduce a class of non-convex penalty functions of the form $P(B) = \sum_{j=1}^p \lambda p(b_j)$, $b_j$ being the $j$-th row of $B$, in multitask regression. We use data depth functions \citep{zuo00} to construct our row-level penalties, which quantify the relative position of $b_j$ with respect to a fixed probability distribution centered at the origin. We approximate this penalty function using local linear approximation, obtain a first level row-sparse estimate, and recover within-row non-zero elements of $B$ through a corrective thresholding of this estimate. When the design matrix is orthogonal and responses independent, the thresholding rule resulting from our proposed penalty has asymptotically optimal minimax risk. Finally we demonstrate the performance of our method relative to some alternatives through a simulation study and microarray data analysis. The supplementary material contains proofs of theoretical results, and additional simulations.

\section{Depth-based regularization}
\label{sec:regression-sec2}

\subsection{Data depth}
Given a data cloud or a probability distribution, a depth function is any real-valued function that measures the outlyingness of a point in feature space with respect to the data or its underlying distribution (figure \ref{fig:fig1} panel a). In order to formalize the notion of depth, we consider as data depth any scalar-valued function $D(x, F_X)$ (where $x \in \mathbb R^p$, and the random variable $X$ has distribution $F_X$) that satisfies the following properties \citep{Liu90}:

\noindent \textit{(P1) Affine invariance}: $D(Ax + b, F_{AX + b}) = D(x, F_X)$ for any $p \times p$ non-singular matrix $A$ and $p \times 1$ vector $b$;

\noindent \textit{(P2) Maximality at center}: When $F_X$ has center of symmetry $\theta$, $D(\theta, F_X) = \sup_{x \in \mathbb R^p} D(x, F_X)$. Here the symmetry can be central, angular or halfspace symmetry;

\noindent \textit{(P3) Monotonicity relative to deepest point}: For any $p \times 1$ vector $x$ and $\alpha \in [0,1]$, $D(x, F_X) \leq D(\theta + a(x - \theta))$;

\noindent \textit{(P4) Vanishing at infinity}: As $\| x \| \rightarrow \infty$, $D(x, F_X) \rightarrow 0$.

Examples of data depth include halfspace depth \citep{tukey75} and projection depth \citep{zuo03}. Data depth has been a popular tool for robust nonparametric and functional inference in the past two decades \citep{jornsten04, ZuoCuiHe04, ZuoCui05, NarisettyNair16}.

\subsection{Motivation}
Given a measure of data depth $D(.,.)$, we define any nonnegative-valued, bounded monotonically decreasing one-to-one transformation on that depth function as an \textit{inverse depth} function, and denote it by $D^-(.,.)$. Some examples of inverse depth transformations include but are not limited to $D^-(x, F_X) := \max_x D(x, F_X) - D(x, F_X)$ and $D^-(x, F_X) := \exp(-D(x, F_X))$. We incorporate inverse depths as row-level penalty functions in (\ref{eqn:penEqn}). Specifically, we estimate $B$ by solving the following constrained optimization problem:
\begin{equation}\label{eqn:eqn02}
\hat B = \argmin_B \left[ \Tr \{ (Y - XB)^T ( Y - XB) \} + \lambda_n \sum_{j=1}^p D^- ( b_j, F) \right]
\end{equation}
We refer to $F$ as the \textit{reference distribution}, and consider it fixed in the estimation process.

In multitask regression, any additive penalty function of the form $P_\lambda (B) = \sum_{j=1}^p \lambda p( b_j)$ regularizes individual rows of the coefficient matrix by providing a control over their distance from the $q$-dimensional origin  through some norm (e.g. the $l_1/l_q$ penalty: \cite{NeghabanWainwright11}), or a combination of norms (e.g. the Adaptive Multi-task Elastic-Net: \cite{ChenEtal12}). Through (\ref{eqn:eqn02}) we generalize this notion by proposing to regularize using the `distance' from a \textit{probability distribution} centered at the origin. Any existing method of norm-based regularization arises as a special case by by using the norm (or combination of norms) as the inverse depth function and taking the degenerate distribution centered at $0$ as $F$ .

Inverse depth functions essentially invert the funnel-shaped contour of the corresponding depth function (panel a of Figure~\ref{fig:fig1}). This immediately results in row-wise nonconvex penalties, where the penalty sharply increases for smaller entries inside the row but is bounded above for large values (see the case for $p=1$ in panel b of Figure~\ref{fig:fig1}). This serves as our motivation of using data depth in regularized multitask regression.

\section{The LARN algorithm}
\label{sec:regression-sec3}

\subsection{Formulation}\label{subsec:subsec31}
The reference distribution $F$ is pivotal in the estimation problem in (\ref{eqn:eqn02}). While we think that there is scope for a significant amount of theoretical analysis on the implications of different choices of $F$ and its potential connections to Bayesian regularized support union recovery in multitask regression \citep{Chen2014}, here we shall work within a simplified setup. Specifically we assume that

\vspace{1em}
\noindent (A1) The distribution $F$ is spherically symmetric.
\vspace{1em}

\noindent This is a fair assumption to make from a frequentist perspective, as we do not possess any extra information about the $q$ responses being different from one another. Since $F$ is spherically symmetric, depth at a point $ b$ becomes a function of $r = \|  b\|_2$ only, due to the affine invariance of $D(.,F)$. In this situation, several depth functions have closed-form expressions: e.g. when $D$ is projection depth and $F$ is a $p$-variate standard normal distribution, $D( b_j, F) = c / (c + r_j); c = \Phi^{-1}(3/4)$ \citep{zuo03}, while for halfspace depth and any known $F$, $D( b_j, F) = 1 - F_1(r_j)$, $F_1$ being any univariate marginal of $F$ (immediate from the definition of halfspace depth) and $r_j = \|  b_j \|_2$. Hence, the computational burden of calculating depths for rows of $ B$ becomes trivial.

\begin{figure}
\begin{center}
\subfigure[]{\epsfxsize=0.31\linewidth \epsfbox{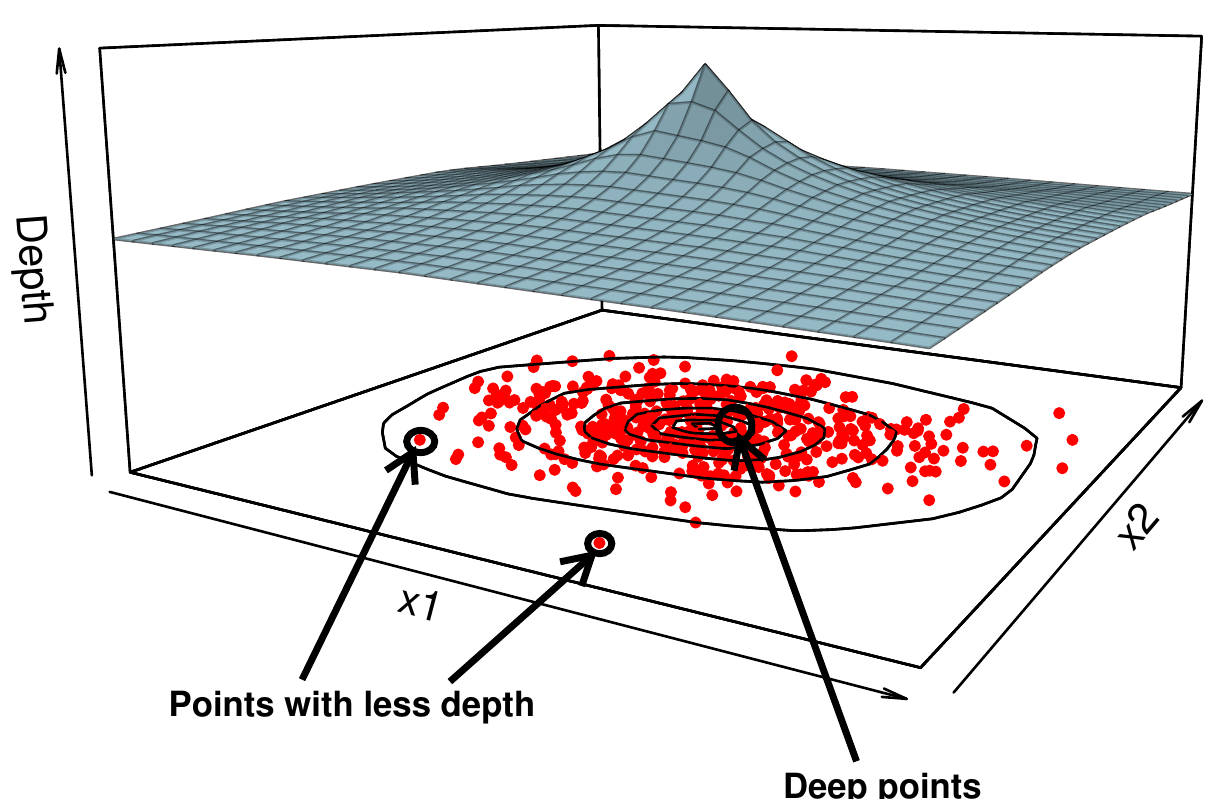}}
\subfigure[]{\epsfxsize=0.31\linewidth \epsfbox{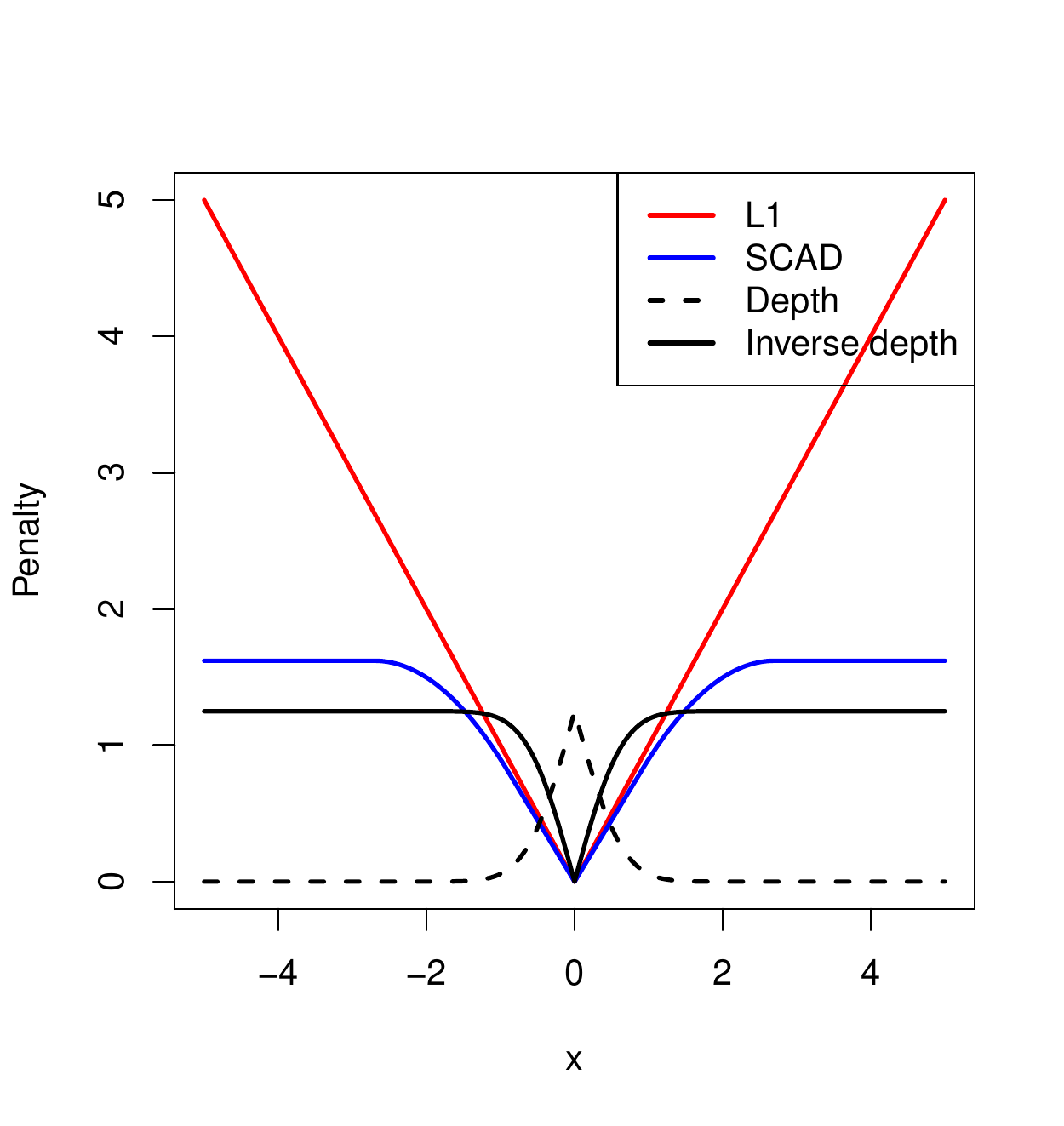}}
\subfigure[]{\epsfxsize=0.31\linewidth \epsfbox{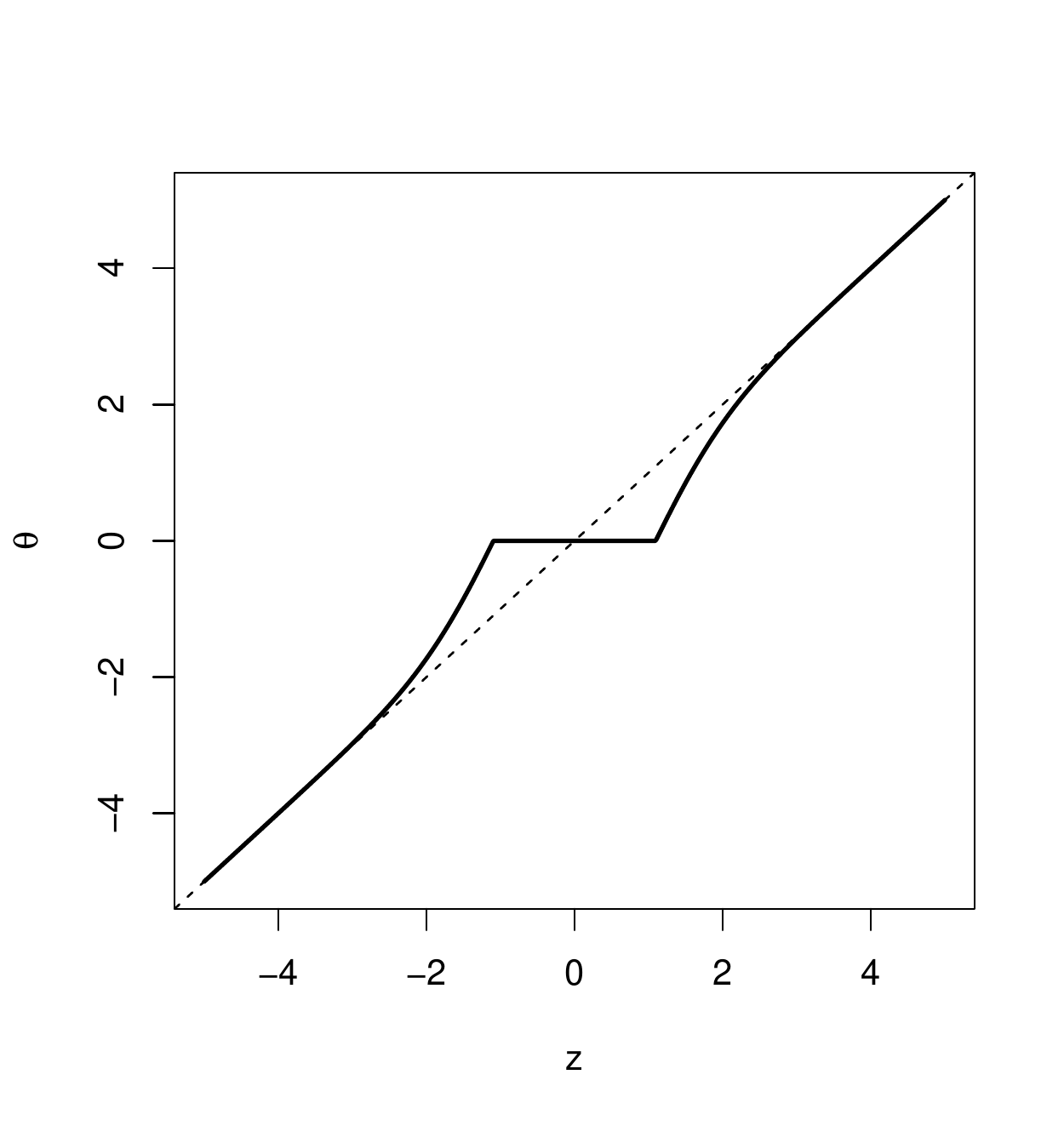}}
\vspace{-1em}
\caption{(a) Contour plot of data depths for 500 points drawn form $\cN_2(0, \diag(2,1))$; (b) Comparison of L1 and SCAD \citep{FanLi01} penalty functions with univariate halfspace depth: inverting the depth function helps obtain the nonconvex shape of the penalty function in the inverse depth; (c) Univariate thresholding rule for the LARN estimate assuming halfspace depth and max definition of inverse depth (see Section~\ref{sec:regression-sec4})}
\label{fig:fig1}
\end{center}
\end{figure}

Because of the way we define inverse depth functions, the above holds for inverse depth functions $D^-(., F)$ as well. Thus we can write that $D^-( b_j, F) = p_F (r_j), r_j = \| b_j \|_2$ for some scalar-valued function $p_F(.)$. Any superscript or subscript in $ B$ or $ b_j$ will be passed accordingly to $r_j$. At this point we make another assumption on $p_F(.)$:

\vspace{1em}
\noindent (A2) The function $p_F(r)$ is concave in $r$, and continuously differentiable at every $r \neq 0$.
\vspace{1em}

\noindent In general depth functions are assumed to have convex contours \citep{MoslerChapter13}, which implies quasi-concavity. Nevertheless, several depth functions adhere to concavity owing to their simplified closed forms for spherical distributions (e.g. halfspace depth and projection depth as stated earlier). Continuous differentiability except at the origin, which is essential for admitting a sparse solution to (\ref{eqn:eqn02}), arises because of the same reason.

Keeping the above setup in mind, we consider the first-order Taylor series approximation of the overall penalty function:
\begin{eqnarray}
\hspace{-1em} P_{\lambda.F} ( B) & = & \lambda \sum_{j=1}^p p_F (r_j) \notag\\
& \simeq & \lambda \sum_{j=1}^p \left[ p_F (r_j^*) + p'_F (r_j^*) ( r_j - r_j^*) \right]\label{eqn:majorizeEqn}
\end{eqnarray}
for any $ B^*$ close to $ B$, and $r_j = \|  b_j \|_2, r_j^* = \|  b_j^* \|_2; j = 1,2,...,p$.

Given a starting solution $ B^*$ close enough to the original coefficient matrix, $P_{\lambda.F} ( B)$ is approximated by its conditional counterpart, say $P_{\lambda.F} ( B |  B^*)$. Following this a penalized maximum likelihood estimate for $ B$ can be obtained using the iterative algorithm below:
\begin{enumerate}
\item Take as starting value $ B^{(0)} = \hat  B
_{\text{LS}} = ( X^T  X)^-  X^T  Y$, i.e. the least square estimate of $ B$, set $k=0$;

\item Calculate the next iterate by solving the penalized likelihood:
\begin{align}\label{eqn:mapEqn}
 \hspace{-3em} B^{(k+1)} = \argmin_ B \left[ \Tr \left\{ ( Y -  X B^{(k)})^T ( Y -  X B^{(k)})\right\} + \lambda \sum_{j=1}^p  p'_F (r_j^{(k)}) r_j \right]
\end{align}

\item Continue until convergence.
\end{enumerate}

Taking $\hat  B_{\text{LS}}$ as a starting value ensures that $\| \hat  B_{\text{LS}} -  B \|_F = O ( n^{-1/2} )$ given the data, hence we get from (\ref{eqn:majorizeEqn}) that
$$
P_{\lambda,F}( B) = P_{\lambda,F} (  B|\hat  B_{\text{LS}} ) + \sum_{j=1}^p o( | r_j - \hat r_{j, \text{LS}} |) = P_{\lambda,F}( B| \hat  B_{\text{LS}} ) + o( n^{-1/2} )
$$
for fixed $p$. This algorithm approximates contours of the nonconvex penalty function using gradient planes at successive iterates, and is a multivariate generalization of the local linear approximation algorithm of \cite{ZouLi08}. We call this the \textit{Local Approximation by Row-wise Norm} (LARN) algorithm.

LARN is a majorize-minimize (MM) algorithm where the actual objective function $Q( B)$ is being majorized by $R ( B |  B^{(k)})$, with
\begin{align*}
\hspace{-2em} Q ( B) &= \Tr \left\{ ( Y -  X B)^T ( Y -  X B)\right\} + P_{\lambda,F} ( B)\\
\hspace{-2em} R ( B |  B^{(k)}) &= \Tr \left\{ ( Y -  X B)^T ( Y -  X B)\right\} 
 + P_{\lambda,F} ( B |  B^{(k)})
\end{align*} 
This is easy to see, because
%
$Q( B) - R( B| B^{(k)})$
 = 
 $\lambda \sum_{j=1}^p \left[ p_F (r_j) - p_F (r_j^*) - p'_F (r_j^*) (r_j - r_j^*) \right]$.
%
And since $p_F(.)$ is concave in its argument, we have $p_F (r_j) \leq p_F (r_j^*) + p'_F (r_j^*) (r_j - r_j^*)$. Thus $Q( B^{(k)}) \leq R( B| B^{(k)})$. Also by definition $Q( B) = R( B^{(k)}| B^{(k)})$.

Now notice that $ B^{(k+1)} = \argmin_ B R( B| B^{(k)})$. Thus $Q( B^{(k+1)}) \leq R( B^{(k+1)}| B^{(k)}) \leq R( B^{(k)}| B^{(k)}) = Q( B^{(k)}) $, i.e. the value of the objective function decreases in each iteration. At this point, we make the following assumption to enforce convergence to a local solution:

\vspace{1em}
\noindent (A3) $Q( B)=Q(M( B))$ only for stationary points of $Q$, where $M$ is the mapping from $ B^{(k)}$ to $ B^{(k+1)}$ defined in \ref{eqn:mapEqn}.
\vspace{1em}

\noindent Since the sequence of penalized losses i.e. $\{ Q( B^{(k)} \}$ is bounded below (by 0) and monotone, it has a limit point, say $\hat B$. Also the mapping $M(.)$ is continuous as $\nabla p_F$ is continuous. Further, we have $Q( B^{(k+1)}) = Q(M( B^{(k)})) \leq Q( B^{(k)})$ which implies $Q(M(\hat B)) = Q(\hat B)$. It follows that $\hat B$ is a stationary point following assumption (A3).

\textit{Remark.} Although the LARN algorithm guarantees convergence to a stationary point, that point may not be a local solution. However, local linear approximation has been found to be effective in approximating nonconvex penalties and obtaining oracle solutions for single-response regression \citep{ZouLi08} and support vector machines \citep{PengWangWu16}. We generalize this concept for the multitask situation.

\subsection{The one-step estimate and its oracle properties}

Due to the row-wise additive structure of our penalty function, supports of each of the iterates $\hat B^{(k)}$ in the LARN algorithm have the same set of singular points as the solution to the original optimization problem, say $\hat B$. Consequently all iterates are capable of producing sparse solutions. In fact, the first iterate itself possesses oracle properties desirable of row-sparse estimates, namely consistent recovery of the non-zero row support of $ B$, as well as of the elements in those rows. 
This is in line with the findings of \cite{ZouLi08} and \cite{FanChen99}.

Given an initial solution $ B^*$, the first LARN iterate, say $\hat B^{(1)}$, is a solution to the optimization problem: 
%
\begin{eqnarray}\label{eqn:OneStepEqn}
\hspace{-2em} \argmin_ B R( B| B^*)  &=& \argmin_ B \left[ \Tr \left\{ ( Y -  X B)^T ( Y -  X B)\right\} + \lambda \sum_{j=1}^p  p'_F (r_j^{(k)}) r_j \right]
\end{eqnarray}
At this point, without loss of generality we assume that the true coefficient matrix $ B$ has the following decomposition: $ B_0 = ( B^T_{01}, 0)^T,  B_1 \in \mathbb R^{p_1 \times q}, 0 < p_1 < p$. Also denote the vectorized (i.e. stacked-column) version of a matrix $ A$ by $\text{vec}( A)$. We are now in a position to to prove oracle properties of the one-step estimator in (\ref{eqn:OneStepEqn}), in the sense that the estimator is able to consistently detect zero rows of $ B$ as well as estimate its non-zero rows as sample size increases:
\begin{Theorem}\label{Thm:OracleThm}

Assume that $ X^T  X/n \rightarrow  C$ for some positive definite matrix $ C$, and $ p'_F(r_j^*) = O( (r_j^* )^{-s})$ for $1 \leq j \leq q, 0 < r_j^* < \delta$ and some $s>0, \delta > 0$. Consider a sequence of tuning parameters $\lambda_n$ such that $\lambda_n / \sqrt n \rightarrow 0$ and $\lambda_n n^{(s-1)/2} \rightarrow \infty$. Then the following holds for the one-step estimate $\hat B^{(1)} = (\hat B^T_{11}, \hat B^T_{10})^T$ (with the component matrices having dimensions $p_1 \times q$ and $p-p_1 \times q$, respectively) as $n \rightarrow \infty$:

\vspace{1em}
\noindent (1) $\ve (\hat B_{10}) \rightarrow 0$ in probability;

\noindent (2) $\sqrt n (\ve (\hat B_{11}) - \ve ( B_{01})) \leadsto \mathcal N_{p_1 q} ( 0, \Sigma \otimes  C_{11}^{-1})$

\vspace{1em}
\noindent where $ C_{11}$ is the first $p_1 \times p_1$ block in $ C$.
\end{Theorem}

The assumption on $ X$ is standard, and ensures uniqueness of the asymptotic covariance matrix of our estimator. The restricted eigenvalue condition, which has been used to establish finite sample error bounds of penalized estimators \citep{NeghabanEtal09} is a stronger version of this. With respect to the general framework of nonconvex penalized $M$-estimation in \cite{LohWainwright15}, $p_F(.)$ satisfies parts (i)-(iv) of Assumption 1 therein, and the conditions of theorem \ref{Thm:OracleThm} adhere to part (v).

\textit{Remark.} The above oracle results depend on the assumption (A1), which simplifies depth as a function of the row-norm. We conjecture that similar oracle properties hold for weaker assumptions. From initial attempts into proving a broader result, we think it requires a more complex approach than the proof of Theorem \ref{Thm:OracleThm}.

\subsection{Recovering sparsity within a row}

The set of variables with non-zero coefficients for each of the $q$ univariate regressions may not be the same, hence recovering non-zero elements \textit{within the rows} is of interest as well. It turns out that consistent recovery at this level can be achieved by simply thresholding elements of the non-zero elements in the one-step estimate obtained in the preceding subsection. \cite{ObozinskiEtal11} have shown that a similar approach recovers within-row supports in multivariate group lasso. The following result formalizes this in our scenario, provided that non-zero signals in $ B$ are large enough:

\begin{Lemma}\label{Thm:RowSupportThm}
Suppose the conditions of theorem \ref{Thm:OracleThm} hold, and additionally all non-zero components of $ B$ have the following lower bound:
$$
| b_{jk} | \geq \sqrt{\frac{16 \log (q p_1) }{C_{min} n}}; \quad 1 \leq j \leq p_1, 1 \leq k \leq q
$$
where $C_{\min} > 0$ is a lower bound for eigenvalues of $ C_{11}$. Also define by $\hat \cS$ the index set of non-zero rows estimated by the LARN algorithm. Then, for some constants $c, c_0 > 0$, the post-thresdolding estimator $T (\hat B^{(1)})$ defined by:
$$
t_{jk} = \begin{cases} 0 & \text{ if } \hat b_{jk}^{(1)} \leq  \sqrt{\frac{8 \log (q|\hat \cS|) }{C_{min} n}}\\
 \hat b_{jk}^{(1)} & \text{ otherwise }
\end{cases}
; \quad j \in \hat \cS, 1 \leq k \leq q
$$
has the same set of non-zero supports within rows as $ B$ with probability greater than $1 - c_0 \exp( - c q \log p_1)$.

\end{Lemma}

\subsection{Computation}
When $ B$ and $ Y -  X  B$ are replaced with their corresponding vectorized versions, the optimization problem in (\ref{eqn:OneStepEqn}) reduces to a weighted group lasso \citep{YangZou15} setup, with group norms corresponding to $l^2$ norms of rows of $ B$ and inverse depths of corresponding rows of the initial estimate $ B^*$ acting as group weights. To compute a solution here, we start from the following lemma, which gives necessary and sufficient conditions for the existence of a solution:
\begin{Lemma}
Given an initial value $ B^*$, a matrix $ B \in \mathbb R^{p \times q}$ is a solution to the optimization problem in (\ref{eqn:OneStepEqn}) if and only if:
\begin{enumerate}
\item $ 2  x_j^T ( Y -  X  B) + \lambda p'_F (r_j^*)  b_j / r_j = 0$ if $ b_j \neq 0$;
\item $ \|  x_j^T ( Y -  X  B) \|_2 \leq \lambda/2 $ if $ b_j = 0$.
\end{enumerate}
\end{Lemma}
This lemma is a modified version of lemma 4.2 in chapter 4 of \cite{BuhlmannBook}, and can be proved in a similar fashion. Following the lemma, we use a block coordinate descent algorithm \citep{LiNanZhu15} to iteratively compute $\hat B^{(1)}$.
%
%
%
%

We use $k$-fold cross-validation to choose the optimal $\lambda$. Additionally, in a sample setup the quantity $C_\text{min}$ in Lemma \ref{Thm:RowSupportThm} is unknown, so we choose a best threshold for within-row sparsity through cross-validation as well. Even though this means that the cross-validation has to be done over a two-dimensional grid, the thresholding step is actually done \textit{after} estimation. Thus for any fixed $\lambda$, only $k$ models need to be calculated. Given a trained model for some value of $\lambda$ we just cycle through the full range of thresholds to record their corresponding cross-validation errors.


\section{Orthogonal design and independent responses}
\label{sec:regression-sec4}

We shed light on the workings of our penalty function by considering the simplified scenario when the predictor matrix $X$ is orthogonal and all responses are independent. Independent responses make minimizing (\ref{eqn:eqn02}) equivalent to solving of $q$ separate nonconvex penalized regression problems, while orthogonal predictors make the LARN estimate equivalent to a collection of coordinate-wise soft thresholding operators.

\subsection{Thresholding rule}

For the univariate thresholding rule, we are dealing with the simplified penalty function $p_F (|b_{jk}|) = D^- (b_{jk}, F)$, where $D^-$ is a inverse depth function based on the univariate depth function $D$. In this case, depth calculation becomes simplified in exactly the same way as in Subsection~\ref{subsec:subsec31}, only $|b_{jk}|$ replacing $\| b_j\|_2$ therein, and $1 \leq k \leq q$.

Following \cite{FanLi01}, a sufficient condition for the minimizer of the penalized least squares loss function
\begin{equation}\label{eqn:eq1}
L(\theta; p_\lambda) = \frac{1}{2} (z - \theta)^2 + p_\lambda(|\theta|)
\end{equation}
to be unbiased when the true parameter value is large is $p'_\lambda (|\theta|)=0$ for large $\theta$. In our formulation, this holds exactly when $F$ has finite support, and approximately otherwise. A necessary condition for sparsity and continuity of the solution is $\min_{\theta \neq 0} |\theta| + p'_\lambda(|\theta|)>0$. We ensure this by making a small assumption about the derivative of $D^-$ (denoted by $D^-_1)$:

\vspace{1em}
\noindent (A4) $\lim_{\theta \rightarrow 0+} D^-_1(\theta,F)> 0$.
\vspace{1em}

Subsequently we get the following thresholding rule as the solution to (\ref{eqn:eq1}):
\begin{eqnarray}\label{eqn:soln_1d}
\hat\theta(F, \lambda) &=& \text{sign} (z) \left[ |z| - \lambda D^-_1(\theta,F) \right]_+ \notag\\
& \simeq & \text{sign} (z) \left[ |z| - \lambda D^-_1(z,F) \right]_+
\end{eqnarray}
The approximation in the second step above is due to \cite{AntoniadisFan01}. A plot of the thresholding function in panel c of Figure~\ref{fig:fig1} demonstrates the unbiasedness and continuity  properties of this estimator.

Thresholding rules due to previously proposed nonconvex penalty functions arise as special cases of our rule. For example, when we use halfspace depth and the max definition of inverse depth, i.e. $D^- ( b, F) = \max_ x D^- ( x, F) - D^- ( b, F)$, the MCP penalty \citep{CHZhang10} corresponds to $D^-_1(\theta,F) = |\theta| \BI_{ |\theta| < \lambda }$, while for the SCAD penalty \citep{FanLi01}:
$$
D^-_1(\theta,F) = \begin{cases} c\lambda & \text{ if } |\theta| < 2 \lambda\\
\frac{c}{a-2} (a \lambda - |\theta|) & \text{ if } 2\lambda \leq |\theta| < a \lambda\\
0 & \text{ if } |\theta| > a \lambda
\end{cases}
$$
with $c = 1/(2\lambda^2(a+2))$.

\subsection{Minimax optimal performance}

In the context of estimating the mean parameters $\mu_i$ of independent and identically distributed observations with normal errors: $z_i = \theta_i + v_i, v_i \sim N(0, 1)$, the minimax risk is $2\log n$ times the ideal risk $R(\text{ideal}) = \sum_{i=1}^n \min (\theta_i^2, 1)$ \citep{DonohoJohnstone94}. A major motivation of using lasso-type penalized estimators in linear regression is that they are able to approximately achieve this risk bound for large sample sizes \citep{DonohoJohnstone94, Zou06}. We now show that our thresholding rule in (\ref{eqn:soln_1d}) also replicates this performance.

\begin{Theorem}\label{thm:minimaxThm}
Suppose the inverse depth function $D^-(.,F)$ is twice continuously differentiable, except at the origin, with first and second derivatives bounded above by $c_1$ and $c_2$ respectively. Then for $\lambda = (\sqrt{.5 \log n}-1)/c_1$, we have
\begin{eqnarray}
\hspace{-2em} R ( \hat\theta(F,\lambda)) & \leq & (2 \log n - 3)
\left[ R(ideal) + \frac{c_1}{p_0(F) (\sqrt{.5 \log n}-1) } \right]
\end{eqnarray}
with $p_0(F) = \lim_{\theta \rightarrow 0+} D^-_1(\theta, F)$.
\end{Theorem}
\noindent Following the theorem, we easily see that for large $n$ the minimax risk of $\hat\theta(F,\lambda)$ approximately achieves the $2 \log n$ multiple bound.

The adaptive lasso \citep{Zou06} guarantees a similar minimax risk bound in single-response regression. This is somewhat expected, given the similar weighted norm structure of the LARN penalty and the adaptive lasso penalty. However, this does \textit{not} hold for all weighted norm penalties: for example the SCAD and MCP penalties do not ensure near-minimax optimal performance because of their non-continuity in the second derivative. In this situation, using inverse depth functions that satisfy all the conditions in the theorem (both halfspace depth and projection depth do because of the simplification in Subsection~\ref{subsec:subsec31}) allows us to go through with the result.

\section{Simulation results}
\label{sec:regression-sec5}

\subsection{Methods and setup}
We use the setup of \cite{RothmanEtal10} in a simulation study to compare the performance of LARN with other relevant methods. Specifically, we use performance metrics calculated after applying the following methods of predictor selection on simulated data for this purpose:

\vspace{1em}
\noindent\textit{LARN}: We use halfspace depth as our chosen depth function and take $D^-( x,F) = \max_ x D( x,F) - D( x,F)$;

\noindent\textit{Thresholded Group Lasso (TGL: \cite{ObozinskiEtal11})}: Performs element-wise thresholding on a row-level group lasso estimator to get final estimate of $ B$. It is a special case of LARN, with weights of all row-norms set to 1;

\noindent\textit{Sparse Group Lasso (SGL: \cite{VincentHansen14})}: This method recovers within row sparsity by considering an $l_1$ penalty over individual elements of $B$ in addition to the $l_1/l_2$ row-level penalties. We use the R package \texttt{lsgl} to fit the model;

\noindent\textit{Separate Lasso}: We train separate lasso models on all response variables with a common tuning parameter.
\vspace{1em}

\noindent For all the methods above, we use 5-fold cross-validation on a 100-length sequence of numbers between $(-2,2)$ as the set of tuning parameters in the respective optimization algorithms. Additionally for LARN and TGL, we use a 100-length sequence between (0, $0.9 \max | \hat B^{(1)} |$) as the set of tuning parameters for within-row thresholding of the first-step estimator $\hat B^{(1)}$.

We generate rows of the model matrix $ X$ as $n=50$ independent draws from $\mathcal{N} (0, \Sigma_X)$, where the positive definite matrix $\Sigma_X$ has a first-order autoregressive (AR(1)) covariance structure, with its $(i,j)^\text{th}$ element given by $0.7^{|i-j|}$. We generate rows of the random error matrix $E$ as independent draws from $\mathcal{N} (0, \Sigma)$: with $\Sigma$ also having an AR(1) structure with correlation parameter $\rho \in \{ 0, 0.5, 0.7, 0.9 \}$. Finally, to generate the coefficient matrix $ B_0$, we obtain the three $p \times q$ matrices: $ W$, whose elements are independent draws from $N(2,1)$; $ K$, which has elements as independent draws from Bernoulli$(0.3)$; and $ Q$ whose rows are made all 0 or all 1 according to $p$ independent draws of another Bernoulli random variable with success probability $0.125$. Following this, we multiply individual elements of these matrices (denoted by $(*)$) to obtain a sparse $ B_0$:
$$
 B_0 =  W *  K *  Q
$$
Notice that the two levels of sparsity we consider: entire row and within-row, are imposed by the matrices $ Q$ and $ K$, respectively.

For a given value of $\rho$, we consider three settings of data dimensions for the simulations: (a) $p=20, q=20$, (b) $p=20, q=60$, (c) $p=60, q=60$ and (d) $p=100, q=60$. Finally we replicate the full simulation 100 times for each set of $(p,q,\rho)$. For brevity, we report only the results for $\rho = 0.7$ here, and provide those for $\rho = 0$, $\rho = 0.5$ and $\rho = 0.9$ in the supplementary material.

\begin{figure}
\begin{center}
\includegraphics[width=\textwidth]{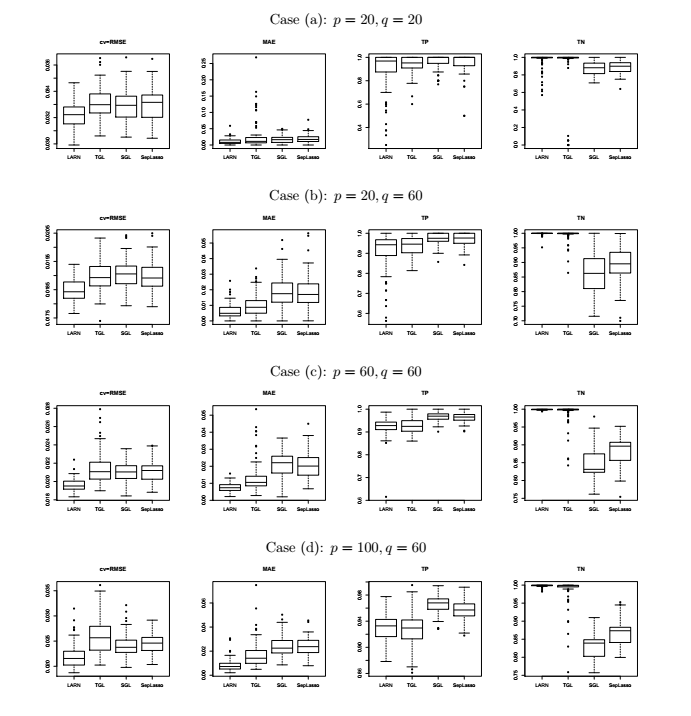}
\end{center}
\caption{Boxplots of evaluation metrics	 for all methods in different $(p,q)$ settings: $\rho = 0.7$.}
\label{fig:simplots}
\end{figure}

\begin{figure}[t]
\begin{center}
\includegraphics[width=.8\textwidth]{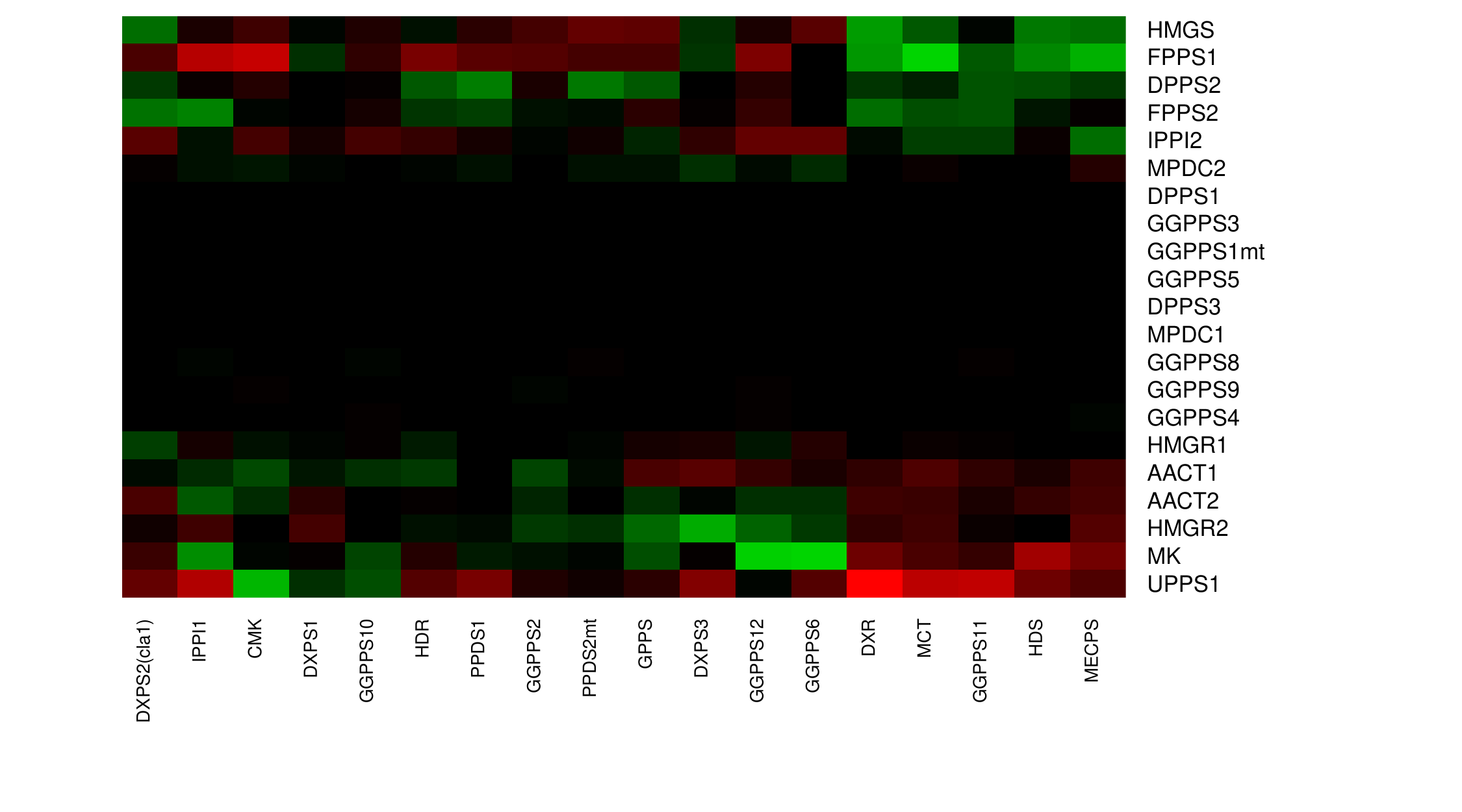}

\caption{Estimated effects of Mevalonate pathway genes (rows) on the activity of non-mevalonate pathway genes (columns) in \textit{A. thaliana}. Green/ red indicate positive/ negative values. Genes were ordered using hierarchical clustering.}
\label{fig:coeffplots}
\end{center}
\end{figure}

\subsection{Evaluation}
To summarize the performance of an estimate matrix $\hat B$ we use the following three performance metrics:

\vspace{1em}
\noindent\textit{Cross-validated Root Mean Squared Error (cv-RMSE)}- Defined as
$$
\text{cv-RMSE} ( \hat  B) = \frac{1}{n q} \sqrt {\sum_{k=1}^K \text{Tr} \left[( Y_k -  X_k \hat  B_{-k} )( Y_k -  X_k \hat  B_{-k} )^T \right]}
$$
for a dataset split into $K$ folds. Here$(Y_k, X_k)$ are the data for samples in the $k$-th fold, and $\hat B_{-k}$ is the estimate obtained from a model trained on samples outside the $k$-th fold;

\noindent\textit{Mean Absolute Error (MAE)}: Defined as the mean absolute value of entries in $\hat B - B_0 $;

\noindent\textit{True Positive Rate (TP)} - The proportion of non-zero entries in $ B_0$ detected as non-zero in $\hat B$;

\noindent\textit{True Negative Rate (TN)} - The proportion of zero entries in $ B_0$ detected as zero in $\hat B$.
\vspace{1em}

A desirable estimate shall have high TP and TN proportions, and low average cv-RMSE and MAE. We summarize the performances of all four methods in Figure~\ref{fig:simplots}. LARN and TGL outperform the other two methods handsomely in all cases. Although their TP and TN performances are similar, LARN estimates perform better in out-of-sample prediction and estimastion of elements in $B$ compared to TGL, owing to lesser cv-RMSE and MAE values. This is expected because the weighted penalties provide asymptotically unbiased estimates for non-zero elements in $B$. Also the performance of TGL varies across all replications by larger amounts compared to LARN in most of the cases considered. Although SGL and SepLasso detect higher number of signals than the thresholded methods, they have high false positive rates. This becomes more severe for higher values of $p$ and $q$. The deterioration of their prediction performance is possibly a result of this.

\section{Gene network data analysis}
\label{sec:regression-sec6}

\begin{figure}[t]
\begin{center}
\includegraphics[width=.8\textwidth]{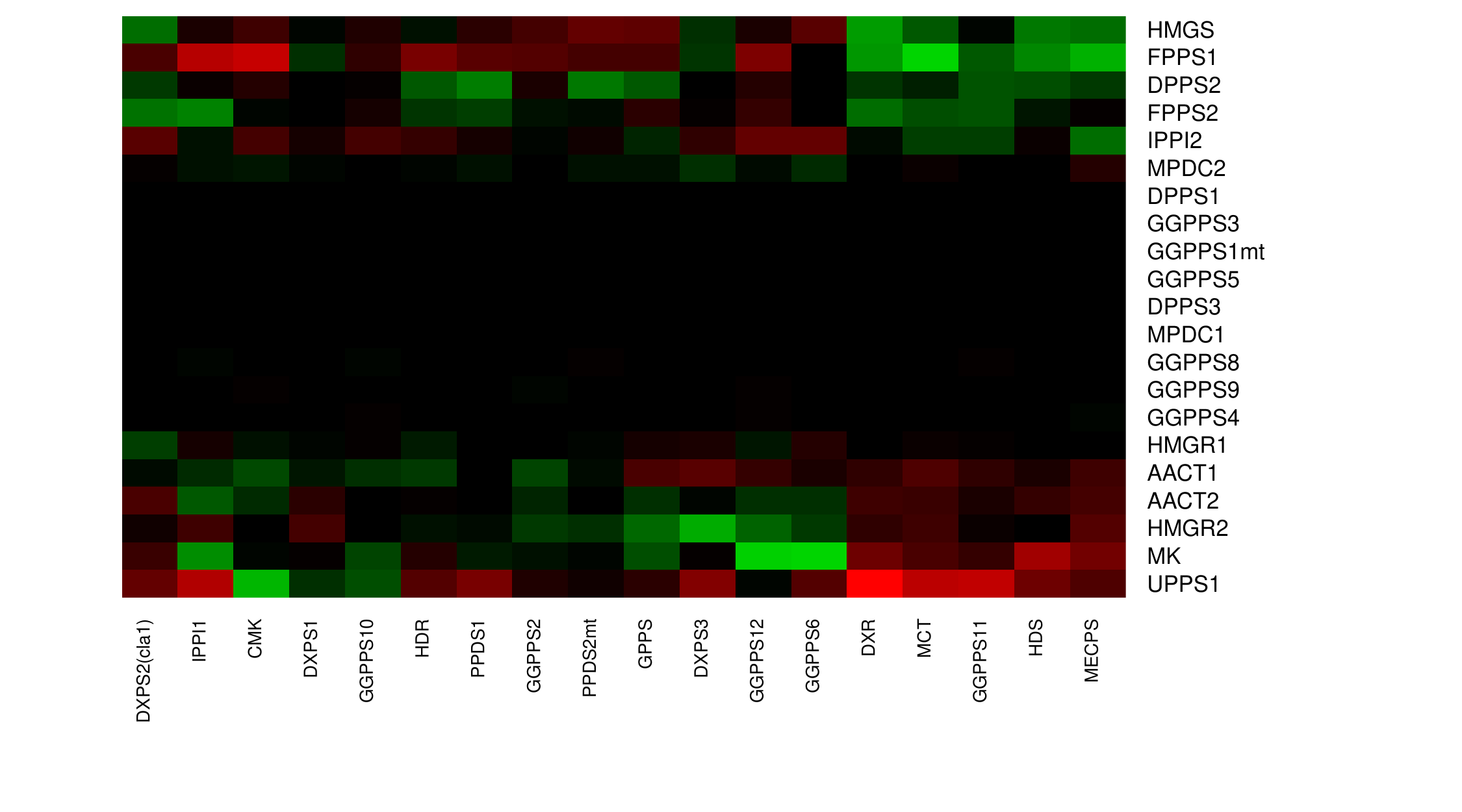}

\caption{Estimated effects of Mevalonate pathway genes (rows) on the activity of non-mevalonate pathway genes (columns) in \textit{A. thaliana}. Green/ red indicate positive/ negative values. Genes were ordered using hierarchical clustering.}
\label{fig:coeffplots}
\end{center}
\end{figure}

We apply the LARN algorithm on a microarray dataset containing expressions of several genes in the flowering plant \textit{Arabidopsis thaliana} \citep{WilleEtal04}. In this dataset, gene expressions are collected from $n=118$ samples, which are plants grown under different experimental conditions. We take the expressions of $q=18$ genes in the non-mevalonate pathway for biosynthesis of isoprenoid compounds, which are key compounds affecting plant metabolism as our multiple responses, and expressions of $p=21$ genes corresponding to the mevalonate pathway as predictors.

Here we want to find out the extent of crosstalk between genes in the two pathways. We apply LARN, and the three methods mentioned before, on the data and evaluate them based on predictive accuracy of 1000 random splits with 100 training samples and 18 test samples: using 5-fold cross-validation to choose optimum values of tuning parameters. LARN has the smallest average RMSE among the four methods, although that comes at the cost of higher number of estimated non-zero elements on average (Table~\ref{table:coefftable}). We summarize the crosstalk between genes in the two pathways by taking elementwise average of the estimated coefficient matrices corresponding to the 1000 random splits. For this average coefficient matrix, we summarize the 10 largest coefficients (in absolute values) in Table~\ref{table:methodtable}, and visualize all coefficients in the table through a heatmap in Figure~\ref{fig:coeffplots}.

\begin{table}[t]
\centering
    \begin{tabular}{l|cccc}
    \hline
    Method                              & LARN        & TGL         & SGL         & SepLasso    \\\hline
    RMSE (x$10^{-2}$)                        & 4.64 (2.1) & 4.74 (2.0) & 4.71 (2.1) & 4.70 (2.1) \\
    Proportion of non-zero coefficients & 0.61 (.008) & 0.66 (.014) & 0.46 (.008) & 0.44 (.019) \\\hline
    \end{tabular}
\caption{Performance of all methods in \textit{A. thaliana} data: mean and standard deviation (in brackets) over 1000 random splits}
\label{table:methodtable}
\end{table}

\begin{table}[t]
\centering
\begin{tabular}{ccc}
 \hline
Coefficient & Mevalonate & Non-Mevalonate\\ 
 &  pathway gene &  pathway gene \\\hline
-0.81 & UPPS1 & DXR \\ 
0.68 & MK & GGPPS6 \\ 
0.67 & FPPS1 & MCT \\ 
0.65 & MK & GGPPS12 \\ 
-0.62 & FPPS1 & CMK \\ 
-0.60 & UPPS1 & GGPPS11 \\ 
-0.59 & UPPS1 & MCT \\ 
0.58 & UPPS1 & CMK \\ 
-0.57 & FPPS1 & IPPI1 \\ 
-0.56 & UPPS1 & IPPI1 \\ 
   \hline
\end{tabular}
\caption{Top 10 between-pathway connections in {\it A. thaliana} data found by LARN}
\label{table:coefftable}
\end{table}

Only 3 genes in the Mevalonate pathway: UPPS1, FPPS1 and MK, control the largest interactions. Among the connections in Table~\ref{table:coefftable}, UPPS1--DXR, MK--GGPPS6, FPPS1--MCT, MK--GGPPS12 and UPPS1--CMK were found previously by \cite{WilleEtal04} (see figures 2 and 3 therein), while the other five are novel. Our other findings also corroborate those obtained by previous studies: for example, the mevalonate pathway genes GGPPS1,3,5,8,9 do not have much effect on the activity of genes in the other pathway \citep{WilleEtal04,LozanoSwirszcz12}.


\section{Conclusion}

Although several nonconvex penalties exist in the literature, the strength of our penalization scheme lies in the significant scope of inference procedures that can rise from the choice of the reference distribution $F$. Our method shares the weakness of all nonconvex penalties: small signals may go undetected or can be estimated in a biased fashion. However the flexibility in choosing $F$ provides enough motivation for further research in fine tuning similar penalization schemes.

\bibliographystyle{plainnat}
\bibliography{depth-regression_101316}

\newpage

\renewcommand{\thesection}{\Alph{section}}
\singlespacing
\begin{huge}
\begin{center}
Supplementary Material for\\

Nonconvex penalized multitask regression using data depth-based penalties
\end{center}
\end{huge}
\doublespacing

\appendix
\section{Proofs}
\label{section:regression-sec8}

\begin{proof}[Proof of Theorem \ref{Thm:OracleThm}]

We shall prove a small lemma before going into the actual proof.

\begin{Lemma}\label{lemma:OracleThmLemma}
For matrices $K \in \mathbb R^{l \times k}, L \in \mathbb R^{l \times m}, M \in \mathbb R^{m \times k}$,
$$
\Tr (K^T L M) = {\ve}^T (K) (I_k \otimes L ) \ve (M)
$$
\end{Lemma}

\begin{proof}[Proof of Lemma \ref{lemma:OracleThmLemma}]
From the property of Kronecker products, $(I_k \otimes L) \ve(M) = \ve (LM)$. The lemma follows since $\Tr (K^T L M) = \ve^T (K) \ve(LM)$.
\end{proof}

Now, suppose $B = B_0 +  U/\sqrt n$, for some $ U \in \mathbb R^{p \times q}$, so that our objective function takes the form
\begin{eqnarray}\label{eqn:OracleThmEqn1}
T_n ( U) &=& \Tr \left[ \left( Y - XB_0 - \frac{1}{\sqrt n}X U \right)^T \left( Y - XB_0 - \frac{1}{\sqrt n}X U \right)\right] \notag\\
&& + \lambda_n \sum_{j=1}^p p'_F ( r_j^*) \left\|  b_{0j} + \frac{ u_j}{\sqrt n} \right\|_2 \notag\\
\Rightarrow T_n ( U) - T_n ({  0}_{p \times q}) &=& \Tr \left[ \frac{1}{n}  U^T X^T X  U - \frac{2}{\sqrt n}  E^T X U\right] \notag\\
&& + \frac{\lambda_n}{\sqrt n} \sum_{j=1}^p p'_F ( r_j^*) \left( \| \sqrt n b_{0j} +  u_j \|_2 - \| \sqrt n b_{0j} \|_2 \right)  \notag\\
&=& \Tr ( V_1 +  V_2) + V_3
\end{eqnarray}

Since $X^TX/n \rightarrow  C$ by assumption, we have $\Tr( V_1) \rightarrow \ve^T ( U)(I_q \otimes  C) \ve( U)$ using Lemma \ref{lemma:OracleThmLemma}. Using the lemma we also get
$$
\Tr( V_2) = \frac{2}{\sqrt n} {\ve}^T( E) (I_q \otimes X) \ve ( U)
$$
Now $\ve( E) \sim \mathcal N_{nq} ({  0}_{n q}, \Sigma \otimes I_q)$, so that $(I_q \otimes X^T) \ve( E)/\sqrt n \leadsto  W \equiv \mathcal N_{pq} ({  0}_{pq}, \Sigma \otimes  C)$ using properties of Kronecker products and Slutsky's theorem.

Let us look at $V_3$ now. Denote by $V_{3j}$ the $j$-th summand of $V_3$. Now there are two scenarios. Firstly, when $ b_{0j} \neq {  0}_q$, we have $p'_F ( r_j^*) \stackrel{P}{\rightarrow} p'_F ( r_{0j})$. Since $\lambda_n / \sqrt n \rightarrow 0$, this implies $V_{3j} \stackrel{P}{\rightarrow} 0$ for any fixed $ u_j$. Secondly, when $ b_{0j} = {  0}_q$, we have
$$
V_{3j} = \lambda_n n^{(s-1)/2}. (\sqrt n r^*_j )^{-s}.\frac{ p'_F ( r_j^*) \|  u_j \|_2 }{ (r^*_j) ^{-s}}
$$
We now have $ b^*_j = O_p(1/\sqrt n)$, and also each term of the gradient vector is $O ((r^*_j)^{-s})$ by assumption. Thus $V_{3j} = O_P( \lambda_n n^{(s-1)/2} \|  u_j \|_2)$. By assumption, $\lambda_n n^{(s-1)/2} \rightarrow \infty$ as $n \rightarrow \infty$, so $V_{3j} \stackrel{P}{\rightarrow} \infty$ unless $ u_j = 0$, in which case $V_{3j} = 0$.

Accumulating all the terms and putting them into \ref{eqn:OracleThmEqn1} we see that
\begin{equation}
T_n( U) - T_n({  0}_{p \times q}) \leadsto
\begin{cases}
\ve^T ( U_1) [ (I_q \otimes  C_{11}) \ve( U_1) - 2  W_1 ] & \text{if }  U_0 = {  0}_{(p-p_1)q}\\
\infty & \text{otherwise}
\end{cases}
\end{equation}
where rows of $ U$ are partitioned into $ U_1$ and $ U_0$ according to the zero and non-zero rows of $B_0$, respectively, and the random variable $ W$ is partitioned into $ W_1$ and $ W_0$ according to zero and non-zero \textit{elements} of $\ve (B_0)$. Applying epiconvergence results of \cite{Geyer94} and \cite{KnightFu00} we now have
\begin{eqnarray}
\ve(\hat U_{1}) &\leadsto & (I_q \otimes  C_{11}^{-1})  W_1\label{eqn:OracleThmProofEqn2}\\
\ve(\hat U_{0}) &\leadsto & {  0}_{(p-p_1)q}\label{eqn:OracleThmProofEqn3}
\end{eqnarray}
where $\hat U = (\hat U_{1}^T, \hat U_{0}^T)^T := \argmin_ U T_n ( U)$.

The second part of the theorem, i.e. asymptotic normality of $\sqrt n (\ve (\hat B_{11}) - \ve(B_{01})) = \hat U_{1}$ follows directly from (\ref{eqn:OracleThmProofEqn2}). It is now sufficient to show that $P( \hat b_j^{(1)} \neq {  0}_q |  b_{0j} = {  0}_q) \rightarrow 0$ to prove the oracle consistency part. For this notice that KKT conditions of the optimization problem for the one-step estimate indicate
\begin{equation}
2  x_j^T (Y - X \hat B^{(1)}) = - \lambda_n p'_F (r_j^*) \frac{ b_j^{(1)}}{r_j^{(1)}} \quad\Rightarrow\quad \frac{2  x_j^T (Y - X \hat B^{(1)})}{\sqrt n} = - \frac{\lambda_n p'_F (r_j^*)}{\sqrt n}. \frac{ b_j^{(1)}}{r_j^{(1)}}
\end{equation}
for any $1 \leq j \leq p$ such that $\hat b_j^{(1)} \neq {  0}_q$. Since $p'_F(r_j^*) = D^-( (r_j^*)^{-s}) = O_P ( \| ( b_{0j} + 1/\sqrt n \|^{-s})$ and $\lambda_n n^{(s-1)/2} \rightarrow \infty$, the right hand side goes to $-\infty$ in probability if $ b_{0j} = {  0}_q$. As for the left-hand side, it can be written as
$$ \frac{2  x_j^T (Y - X \hat B^{(1)})}{\sqrt n} = \frac{2  x_j^T X .\sqrt n (B_0 - \hat B^{(1)})}{n} + \frac{2  x_j^T  E}{\sqrt n} = \frac{2  x_j^T X \hat U_n}{n} + \frac{2  x_j^T  E}{\sqrt n}
$$
Our previous derivations show that vectorized versions of $\hat U_n$ and $ E$ have asymptotic and exact multivariate normal distributions, respectively. Hence
$$
\BP \left[ \hat b_j^{(1)} \neq {  0}_q |  b_{0j} = {  0}_q \right] \leq P \left[ 2  x_j^T (Y - X \hat B^{(1)} ) = - \lambda_n p'_F (r_j^*) \frac{ b_j^{(1)}}{r_j^{(1)}} \right] \rightarrow 0
$$
\end{proof}

\begin{proof}[Proof of Lemma \ref{Thm:RowSupportThm}]
See the proof of corollary 2 of \cite{ObozinskiEtal11} in Appendix A therein. Our proof follows the same steps, only replacing $\Sigma_{SS}$ with $\Sigma \otimes  C_{11}$.

\end{proof}

\begin{proof}[Proof of Theorem \ref{thm:minimaxThm}]
We broadly proceed in a similar fashion as the proof of Theorem 3 in \cite{Zou06}. As a first step, we decompose the mean squared error:
\begin{eqnarray*}
E[ \hat\theta(F,\lambda) - \theta]^2 &=& E[ \hat\theta(F,\lambda) - z]^2 + E(z - \theta)^2 + 2 E [\hat\theta(F,\lambda) (z-\theta)] - 2E [z(z-\theta)]\\
&=& E[ \hat\theta(F,\lambda) - z]^2 + E \left[ \frac{d\hat\theta(F,\lambda)}{dz}\right] - 1
\end{eqnarray*}
by applying Stein's lemma \citep{Stein81}. We now use Theorem 1 of \cite{AntoniadisFan01} to approximate $\hat\theta(F,\lambda)$ in terms of $y$ only. By part 2 of the theorem,
\begin{equation}
\hat\theta(F,\lambda) = \begin{cases}
0 \quad & \text{if } |z| \leq \lambda p_0(F)\\
z - \text{sign}(z). \lambda D^-_1(\hat\theta(F,\lambda), F) & \text{if }|z| > \lambda p_0(F)
\end{cases}
\end{equation}
Moreover, applying part 5 of the theorem,
\begin{equation}
\hat\theta(F,\lambda) = z - \text{sign}(z).\lambda D^-_1(z, F) + o(D^-_1(z, F))
\end{equation}
for $|z| > \lambda p_0(F)$. Thus we get
\begin{equation}
[ \hat\theta(F,\lambda) - z]^2 = \begin{cases}
z^2 & \text{if } |z| \leq \lambda p_0(F)\\
\lambda^2 D^-_1(z,F)^2 + k_1(|z|) & \text{if } |z| > \lambda p_0(F)
\end{cases}
\end{equation}
and
\begin{equation}
\frac{d\hat\theta(F,\lambda)}{dz} = \begin{cases}
0 & \text{if } |z| \leq \lambda p_0(F)\\
1 +  \lambda D^-_2(z,F) + k_1'(|z|) & \text{if } |z| > \lambda p_0(F)
\end{cases}
\end{equation}
where $k_1(|z|) = o(|z|)$, and $D^-_2(z,F) = d^2D^-(z,F)/dz^2$. Thus
\begin{eqnarray}\label{eqn:MinimaxProofEqn1}
E [ \hat\theta(F,\lambda) - \theta]^2 &=& E [z^2 \BI_{|z| \leq \lambda p_0(F)}] + E \left[ \left(\lambda^2 D^-_1(|z|, F)^2 + 2 \lambda D^-_2(|z|,F) + 2 + \right.\right. \notag\\
&& \left.\left. k_1(|z|) + k_1'(|z|) \right) \BI_{|z| > \lambda p_0(F)} \right] - 1
\end{eqnarray}
Now
\begin{eqnarray*}
k_1(|z|) &=& \lambda^2 \left[ D^-_1(z,F)^2 - D^-_1(\hat\theta(F,\lambda), F)^2 \right] \quad \leq \quad \lambda^2 c_1^2, \text{ and}\\
| k_1'(|z|) | &=& \lambda \left| D^-_2(z,F) - \frac{d D^-_1(\hat\theta(F,\lambda), F)}{dz} \right| \quad \leq \quad 2\lambda c_2
\end{eqnarray*}
Substituting these in (\ref{eqn:MinimaxProofEqn1}) above we get
\begin{eqnarray}\label{eqn:MinimaxProofEqn2}
E [ \hat\theta(F,\lambda) - \theta]^2 & \leq & \lambda^2 p_0(F)^2 P [|z| \leq \lambda p_0(F)] + E \left[ \left(\lambda^2 f^2(|z|) + 2 \lambda D^-_2(z, F) \right) BI_{|z| > \lambda p_0(F)} \right] \notag\\
&& + \lambda^2 c_1^2 + 2\lambda c_2 + 1 \notag\\
& \leq & 2\lambda^2 c_1^2 + 4 \lambda c_2 + 1 \notag\\
& \leq & 4\lambda^2 c_1^2 + 8 \lambda c_2 + 1
\end{eqnarray}

Adding and subtracting $z^2 \BI_{|z| > \lambda p_0(F)}$ to the first and second summands of (\ref{eqn:MinimaxProofEqn1}) above, we also have
\begin{eqnarray}
E [ \hat\theta(F,\lambda) - \theta]^2 &=& Ez^2 + E \left[ \left(\lambda^2 D^-_1(z,F)^2 + 2 \lambda D^-_2(z,F) + 2 - y^2 + \lambda^2 c_1^2 \right.\right. \notag\\
& & \left.\left. + 2\lambda c_2 \right) \BI_{|z| > \lambda p_0(F)} \right] - 1\notag\\
& \leq & (2 \lambda^2 c_1^2 + 4 \lambda c_2) P [ |z| > \lambda p_0(F) ] + \theta^2
\end{eqnarray}
Following \cite{Zou06}, $P[|z| > \lambda p_0(F)] \leq 2q (\lambda p_0(F)) + 2\theta^2$, with $q(x) = \exp[-x^2/2]/(\sqrt{2\pi} x)$. Thus
\begin{eqnarray}
E [ \hat\theta(F,\lambda) - \theta]^2 &\leq & 2 (2 \lambda^2 c_1^2 + 4 \lambda c_2) [q (\lambda p_0(F)) + \theta^2] + \theta^2\notag\\
& \leq & (4\lambda^2 c_1^2 + 8 \lambda c_2 + 1)[q (\lambda p_0(F)) + \theta^2] 
\end{eqnarray}
Combining this with (\ref{eqn:MinimaxProofEqn2}) we get
\begin{equation}
E [ \hat\theta(F,\lambda) - \theta]^2 \leq [ 4(\lambda c_1 + 1)^2 - 3][q (\lambda p_0(F)) + \min(\theta^2,1)] 
\end{equation}
assuming without loss of generality that $c_1 \geq c_2$. Since $R(\text{ideal}) = \min(\theta^2,1)$ and $q(x) \leq (\sqrt{2\pi} x)^{-1} < 1/x$, we have the needed.
\end{proof}

\section{Additional simulations}
\begin{figure}
\begin{center}
\includegraphics[width=\textwidth]{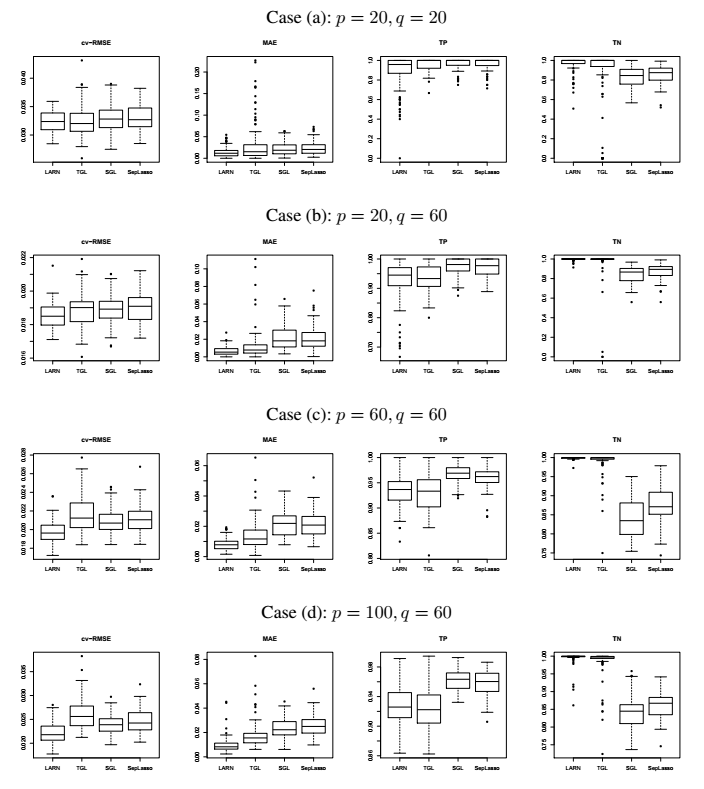}
\end{center}
\caption{Boxplots of evaluation metrics	 for all methods in different $(p,q)$ settings: $\rho = 0$.}
\label{fig:simplots2}
\end{figure}

\begin{figure}
\begin{center}
\includegraphics[width=\textwidth]{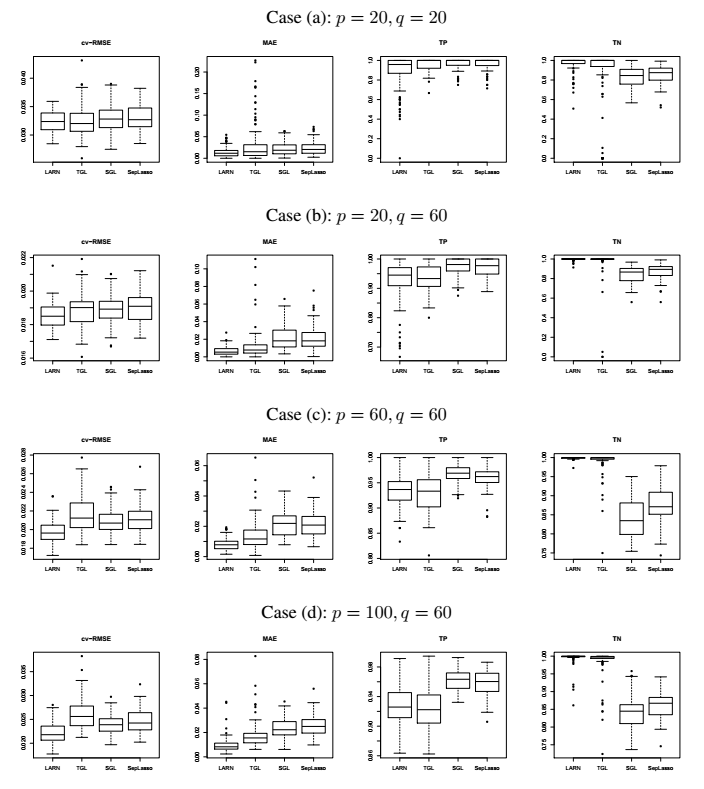}
\end{center}
\caption{Boxplots of evaluation metrics	 for all methods in different $(p,q)$ settings: $\rho = 0.5$.}
\label{fig:simplots3}
\end{figure}

\begin{figure}
\begin{center}
\includegraphics[width=\textwidth]{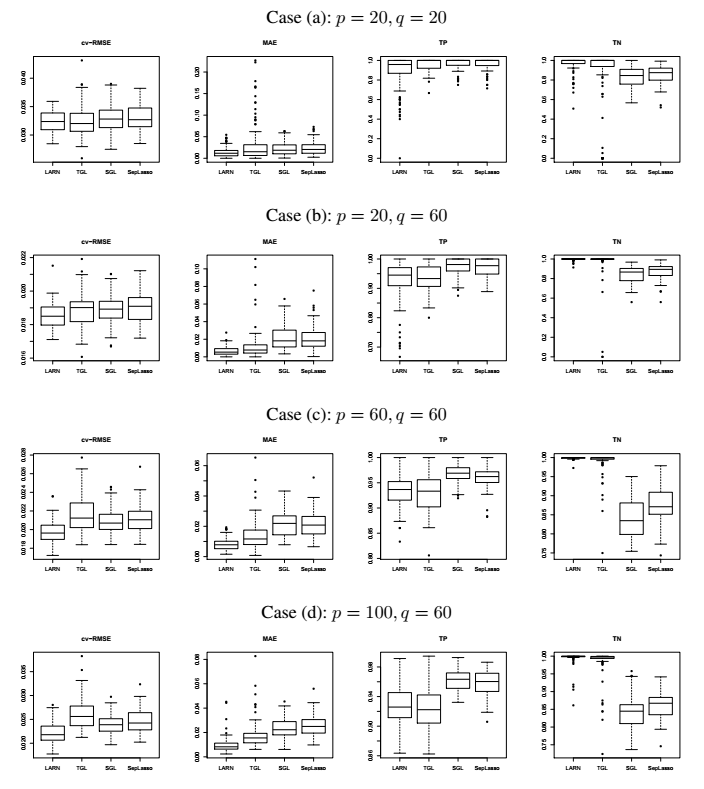}
\end{center}
\caption{Boxplots of evaluation metrics	 for all methods in different $(p,q)$ settings: $\rho = 0.9$.}
\label{fig:simplots4}
\end{figure}

We present the simulation results corresponding to $\rho=0$, $\rho=.5$ and $\rho=.9$ in Figures \ref{fig:simplots2}, \ref{fig:simplots3} and \ref{fig:simplots4}, respectively. The results are similar to the case of $\rho=.7$ presented in the main paper. LARN has the lowest MAE in all cases, and the lowest cv-RMSE in all but one (Case (a) for $\rho=0$) cases.

\end{document}